\documentclass[prl,twocolumn,reprint,showpacs,preprintnumbers,amsmath,amsthm,amssymb,amsfont]{revtex4-1}
\pdfoutput=1
\usepackage{times}
\usepackage{graphicx}
\usepackage{dcolumn}
\usepackage{bm}
\usepackage{color}
\usepackage{subfigure}

\providecommand{\angs}{\rm \AA}

\begin{document}

\title{Origin of lowered magnetic moments in epitaxially strained thin films of multiferroic Bi$_2$FeCrO$_6$}
\author
{Paresh C. Rout$^{(1)}$, Aditya Putatunda$^{(3)}$and Varadharajan Srinivasan$^{(1,2)}$}
\affiliation{(1) Department of Physics, Indian Institute of Science Education and Research Bhopal, Bhopal 462 066, India}
\affiliation{(2) Department of Chemistry, Indian Institute of Science Education and Research Bhopal, Bhopal 462 066, India}
\affiliation{(3) Department of Physics, National Institute of Science Education and Research Bhubhaneswar, Bhubhaneswar 751 005,India}

\begin{abstract}
We have investigated the effect of epitaxial strain on the magnetic properties and $B$-site cation ordering in multiferroic Bi$_2$FeCrO$_6$ (001) thin films using a density-functional theory approach. We find that in thin films with rock-salt ordering of Fe and Cr the ground state is characterised by C-type anti-ferromagnetic (AFM) order. This is in contrast to the bulk form of the material which was predicted to be a ferrimagnet with G-type AFM order. Furthermore, the cation ordered thin-films undergo a transition with epitaxial strain from C to A-type AFM order. Other magnetic orders appear as thermally accessible excited states. We also find that $B$-site cation disordered structures are more stable in coherent epitaxial strains thereby explaining the lowered magnetic moments observed in these samples at room temperature. Strain varies both the sign as well as strength of the Fe-Cr superexchange coupling resulting in a very interesting phase diagram for Bi$_2$FeCrO$_6$ thin films.
\end{abstract}


\maketitle

Multiferroic materials are characterised by coexistence of ferroic orders such as ferroelectricity, ferromagnetism and/or ferroelasticity along with a coupling of at least two of these orders. These properties have enormous technological implications~\cite{spaldin,catalan}. Hence, much attention has recently been directed towards understanding the origins of multiferroicity and the design of multiferroic materials~\cite{eeren, cheong, ramesh, dagotto, greenblatt}. 

Presently there are only a few examples of candidate multiferroics most of which exhibit weak magneto-electric couplings, particularly at room temperature. The desire for a multiferroic with significant values of both polarisation and magnetisation at room temperature motivated the {\it ab initio} design of the double-perovskite  Bi$_2$FeCrO$_6$ (BFCO) derived from the parent multiferroic BiFeO$_3$ (BFO)~\cite{baettig,spladin}. Both compounds posses rhombohedral distorted structure in bulk form with space group {\it R}3{\it c}. The structure consists of Fe and Cr cations occupying the $B$-sites on alternate (111) layers (rock-salt or double-perovskite ordering). A G-type anti-ferromagnetic (AFM) order  was predicted in bulk BFCO  with Fe and Cr sub-lattices having opposite spins. This mirrors the ordering seen in the parent compound. The source of ferroelectric property of both BFO and BFCO is the $6s^{2}$ lone pair on Bi whereas the ferromagnetism arises from the superexchange interaction of magnetic ion pairs Fe$(d^5)$-Fe${(d^5)}$  and Fe${(d^5)}$-Cr$(d^{3}$), respectively~\cite{baettig}. In BFCO this implies a bulk magnetic moment of $2\mu_B$ per formula unit making it a ferrimagnet. 

Experimental efforts to realise BFCO have resulted in several detailed studies of epitaxially grown BFCO on oxide substrates~\cite{neh2,neh1,neh3,khare}. These studies have confirmed the existence of a room temperature magnetisation as well as large polarisation. However, the magnetic moments measured are generally up to an order of magnitude lower than the theoretically predicted value in most films whereas the polarisation remains relatively unaffected. The electric and magnetic properties of BFCO films have been studied as function of epitaxial strain and orientation. It has emerged that under most conditions the double-perovskite structure predicted by baettig {et al.}~\cite{baettig} is only metastable in the films which instead have the transition metal ions disordered over the $B$-sites. This disordering of Fe and Cr  ions has been suggested to be responsible for the low magnetisation seen due to local cancelation of moments. Recently, Khare {\it et al.}~\cite{khare} observed that the magnetisation in epitaxially grown La-doped BFCO (001) thin films at room temperature is not only significantly reduced ($<0.3 \mu_B$/f.u.) compared to the bulk value but increases in magnitude from tensile to compressive strains. The connection between the magnetic moments in the films, epitaxial strain and the $B$-site cation ordering is as yet unclear and can be best provided by first-principles simulations. However, no such studies presently exist in the literature.
\begin{figure*}[t!]
\centering
\subfigure[]
{
\includegraphics[width=0.45\textwidth, clip=]{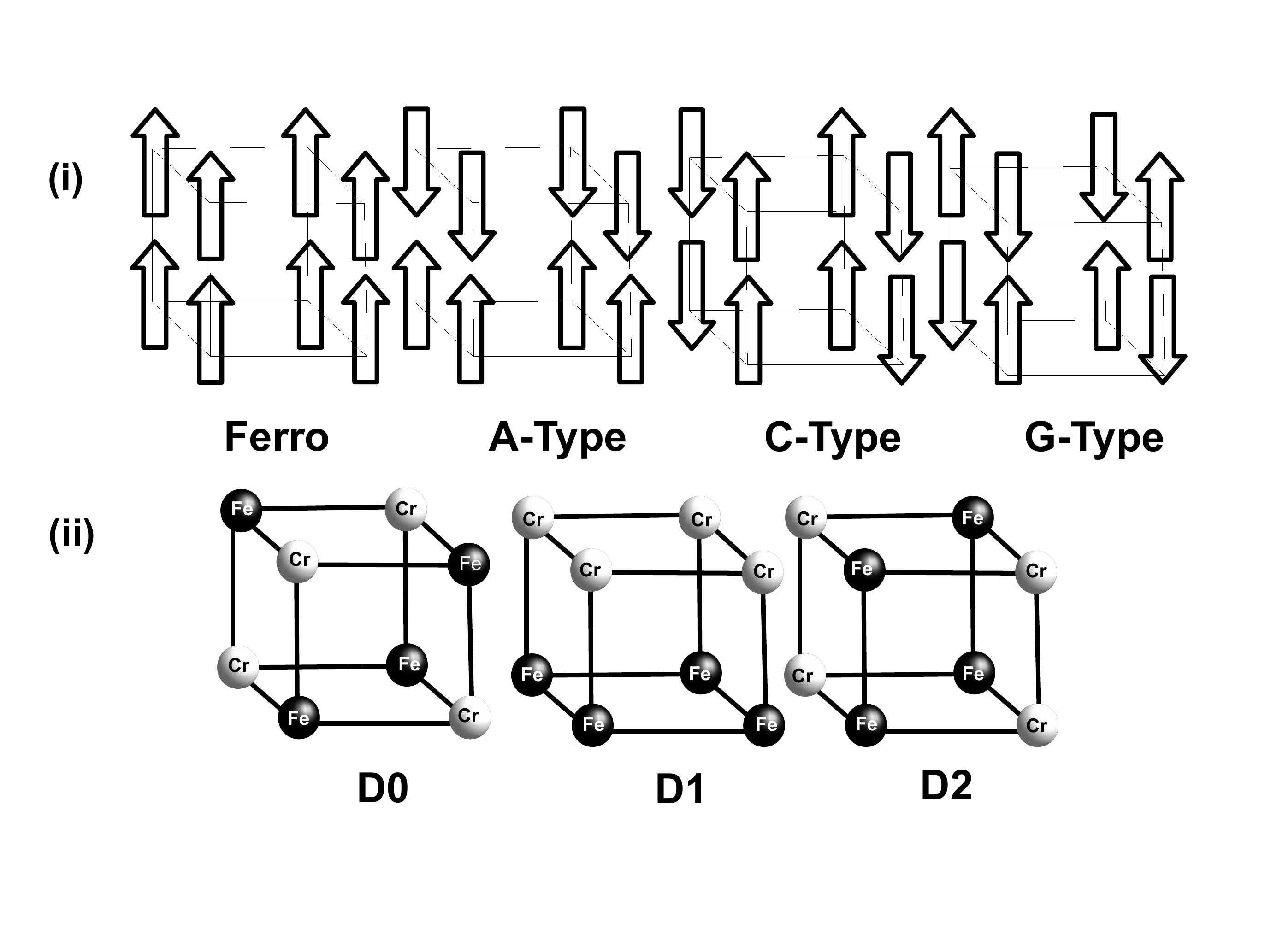}
\label{fig1:subfig1}
}
\subfigure[]
{
\includegraphics[trim= 20 50 65 80, clip, width=0.5\textwidth]{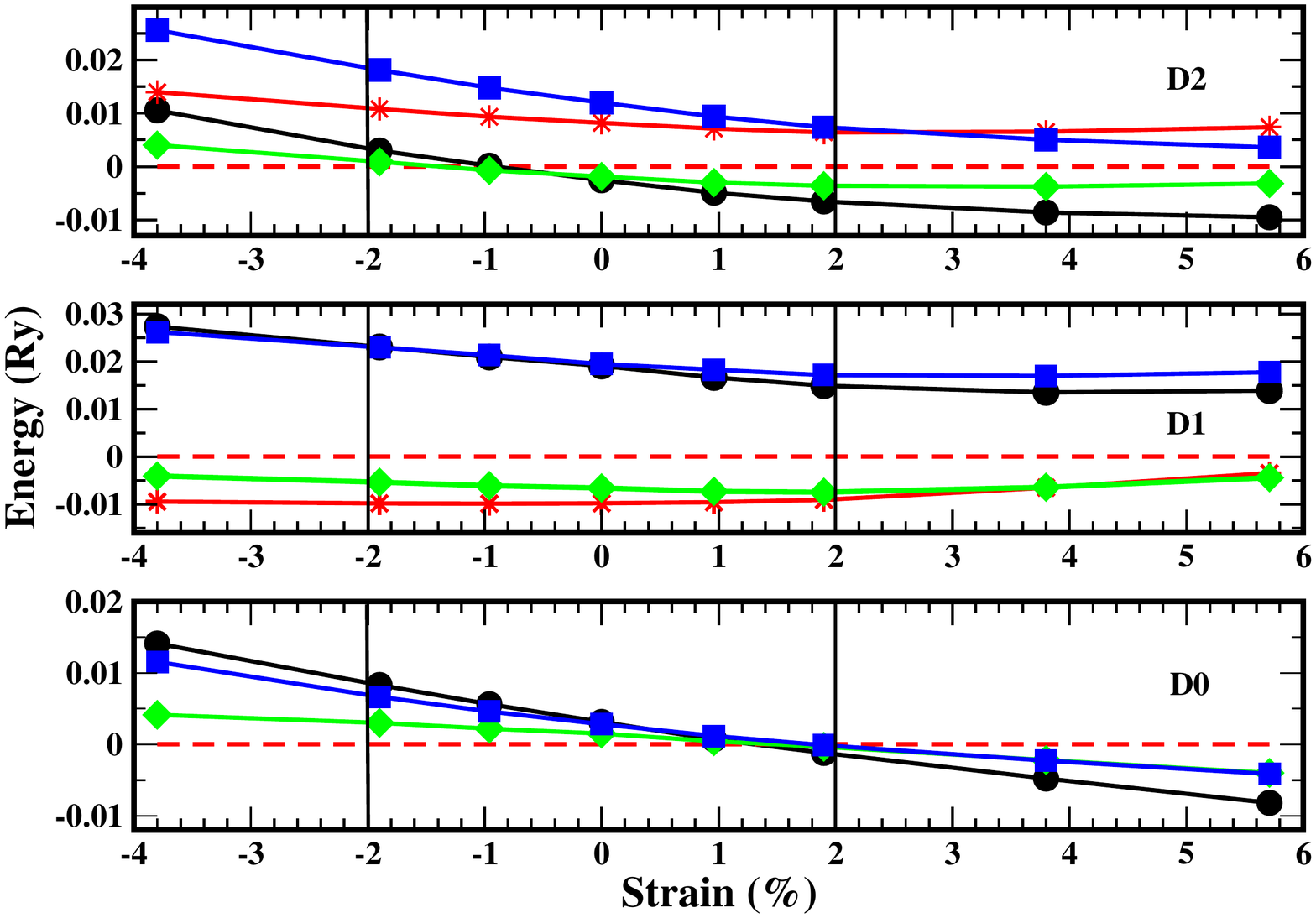}
\label{fig1:subfig2}
}
\caption{(a) Different types of super cells considered in the calculations differing in : (i) magnetic ordering, with up and down arrows representing up and down spins, respectively, and  (ii) cation ordering giving D0 (double perovskite), D1 and D2 structures., (b) Energy vs strain for all magnetic and chemically ordered structures of BFCO. Energies are plotted relative to the D0 C-type structures (broken line) for the D0 structure (bottom panel), D1 structure (middle panel) and the D2 structure (top panel). In all panels the following symbols are used for the magnetic orders : asterisks for C-type, diamonds for G-type, circles for A-type and squares are for ferromagnetic order. \label{fig1:subfig1}}
\end{figure*}

In this study, we investigate the origins of the low magnetic moments and the strain dependence of the moments in epitaxially grown BFCO (001) thin films by employing first-principles density-functional theory (DFT) simulations. By considering a variety of magnetic and $B$-site cation orderings in the thin films we establish that contrary to the bulk the ground-state, AFM order in these films is not  of the G-type. Throughout coherent strains we observe the C-type AFM ordering to be the most stable while other orders appear as thermally accessible excited states. Furthermore, we also find that structures with anti-site defects (see Fig.\ref{fig1:subfig1}) are more stable than the rock-salt ordered structure through all coherent strains. For the rock-salt ordered structure we find that epitaxial strain not only induces a transition in the magnetic order from C-type AFM  to A-type AFM but also lowers the gap to the G-type AFM and ferromagnetic (FM) orders. Since the latter two are associated with non-zero moments this means a potential enhancement in the magnetic moment with strain at room temperature. The drastic variation of the energies of these magnetic orders can be understood by the variation of the in-plane and out-of-plane exchange coupling constants. This is possible because the superexchange interaction in BFCO involves a $d^5-d^3$ pair which, depending on the Fe-O-Cr bond angle, can be either FM or AFM in nature~\cite{moskvin,LFCO1,Pickett}. We have confirmed this variation by extracting the coupling constants from supercell calculations at different strains. We have also used the coupling constants thus extracted in a finite temperature 3D Monte Carlo simulation to calculate the layer distribution of the thin films over various magnetic orders. Thus we can estimate the  magnetisation of the sample at room temperature under coherent epitaxial strain averaged over different cation orderings (Fig.\ref{fig1:subfig1}b). We find the estimated magnetisation to be of similar magnitudes as measured in experiments~\cite{khare}. However, our results indicate almost no strain-dependence of the moments at least in the coherent strain region. Any dependence must either result from the differing proportions of cation ordering in the experimental samples or from spin canting which has not been accounted for in this work.

Experimentally, it was found that BFCO (001) films grow with a tetragonal ($Pbnm$) structure~\cite{nehpol,khare} under coherent strains. Thus, for our calculations we constructed a 20-atom $\sqrt{2}\times\sqrt{2}\times2$ tetragonal supercell (inset in Fig.\ref{fig1:subfig1}a), starting from a  simple cubic double perovskite structure, to allow for appropriate magnetic ordering of ions along (111) direction. Using the experimental in-plane pseudo-cubic lattice parameter $a _{cub}=3.93\angs$~~\cite{khare,nache} as reference we generated structures mimicking the epitaxially-strained films by varying the in-plane lattice parameter. The in-plane lattice parameters for our supercell were set to $\bar{a}=\bar{b}=\sqrt{2}\times a_{cub}$ while the $c$ parameter was relaxed for each in-plane strain. The epitaxial strain was defined as $\epsilon=(\bar{a} -\bar{a}_{ref})/\bar{a}_{ref}$ where $ \bar{a}_{ref} $ is the unstrained lattice parameter. 

Our calculations employed a spin-polarized GGA+$U$~\cite{anisimov,matteo,wentzco} approach using the Perdew-Burke-Ernzerhof~\cite{pbe1,pbe2} exchange-correlation functional within the framework of the Quantum-ESPRESSO code~\cite{paulo}. We chose $U=3.0$ eV which reproduced well the band structure obtained in previous LSDA+$U$ calculations on BFCO~\cite{baettig,Ghosez}. For further details we refer the reader to the supplementary information (SI).

We first investigated the effect of epitaxial strain on the G-type AFM ordered BFCO since this was the order predicted for the bulk. For each value of the strain $\epsilon$ the atomic positions as well as the out-of-plane lattice parameter were relaxed yielding a value $c_{min}(\epsilon)$ that minimises the total energy. The calculated $c_{min}(\epsilon)$ (Fig.\ref{figureS1a} in SI) are within 1.0\% of the experimentally reported values~\cite{khare} for similar strains. Similar agreements have been obtained in previous first-principles studies on epitaxially strained films~\cite{dobin} giving us confidence in the method employed presently. We note here that in the experiments~\cite{khare} coherently strained films formed only in a window of 4\% about the reference strain. Hence our results would be most relevant in this strain window. Our calculations also extended beyond this window in order to clearly illustrate the effect strain has on the various interactions in the system. %However, in all the results below we have clearly indicated the experimentally relevant range. 

The average magnetic moment per formula unit (f.u.) was calculated at each $c_{min}$. But the spontaneous magnetisation remained flat throughout the coherent strain region at the bulk value of 2$\mu_B$. This trend clearly disagrees with the observation of lowered moments in experiments~\cite{khare} and motivated us to consider the possibility of different ground state magnetic and cation ordering. 

We considered three possible types of AFM ordered structures - A, C and G-types - and a FM ordered structure (see Fig.\ref{fig1:subfig1}(b)). In each of these magnetic orders we further allowed for cation disorder  by considering two different arrangements of the Fe/Cr ions in the supercell (anti-site defects) besides the double-perovskite structure, D0. These two different types of  defect structures (Fig.\ref{fig1:subfig1}(b)) are referred to as D1 and D2 structures below. These defects differ in the occupation of the perovskite $B$-sites in the supercell. Note that only a few of these structures result in a non-zero magnetic moment, {\it viz.} G-type in D0, A-type in D1 and C-type in D2. Needless to say, in all structures the FM arrangement leads to a non-zero magnetic moment as well (8 $\mu_B$/f.u.). Experimentally grown thin films may present lower concentration of anti-site defects than those considered here, implying the need for larger super-cells in calculations. However, in the spirit of keeping the analysis simple we restrict our studies to a very limited set of disordered structures within the 20-atom super-cell.

Taking the D0 (double-perovskite) supercell we considered each one of the magnetic orders in Fig.\ref{fig1:subfig1}(b) and carried out relaxations similar to the one described above. The minimised energy at each epitaxial strain is plotted against strain in Fig.\ref{fig1:subfig2}. Interestingly, the C-type AFM order emerged as the ground state from -4\% strain upto 1.5\% strain. This is in contrast to what is anticipated from bulk BFCO. The G-type and A-type AFM, and the FM orders appear as excitations from this phase. In particular, the G-type is very close in energy to the ground-state order being easily thermally accessible. From 1.5\% to 4\% strain the A-type AFM order has the lowest in energy with the G-type AFM and FM ordered excitations being nearly degenerate. This clearly indicates the possibility of a strain-induced transition in the magnetic order in BFCO thin films. Compressive strains seem to prefer the C-type AFM order while tensile strains induce A-type AFM order. These results are interpreted in terms of the super-exchange coupling constants below.

Similar calculations were also carried out for D1 and D2 structures. The energy versus strain plots of all structures (Fig.\ref{fig1:subfig2}) suggest an interesting phase diagram of different possible orders. It was found that under all
strains the D1 structure in the C-type AFM ordering was the most stable. The ground-state order in the D1 structure yielded a zero magnetic moment and dominates the phase diagram in the experimental range of strains. This could explain the lowered observed moments in the films compared to the bulk value. Interestingly, the ground states for all BFCO structures have a zero moment indicating that the experimentally observed moments arise primarily from thermal excitations. For the analysis below we focus only on the regions of strains $\vert\epsilon\vert<2$ as coherently strained films form only in this window~\cite{khare}. 

\begin{figure}[t!]
\centering
\subfigure[]
{
\includegraphics[trim= 20 30 65 80, width=0.45\textwidth, clip=]{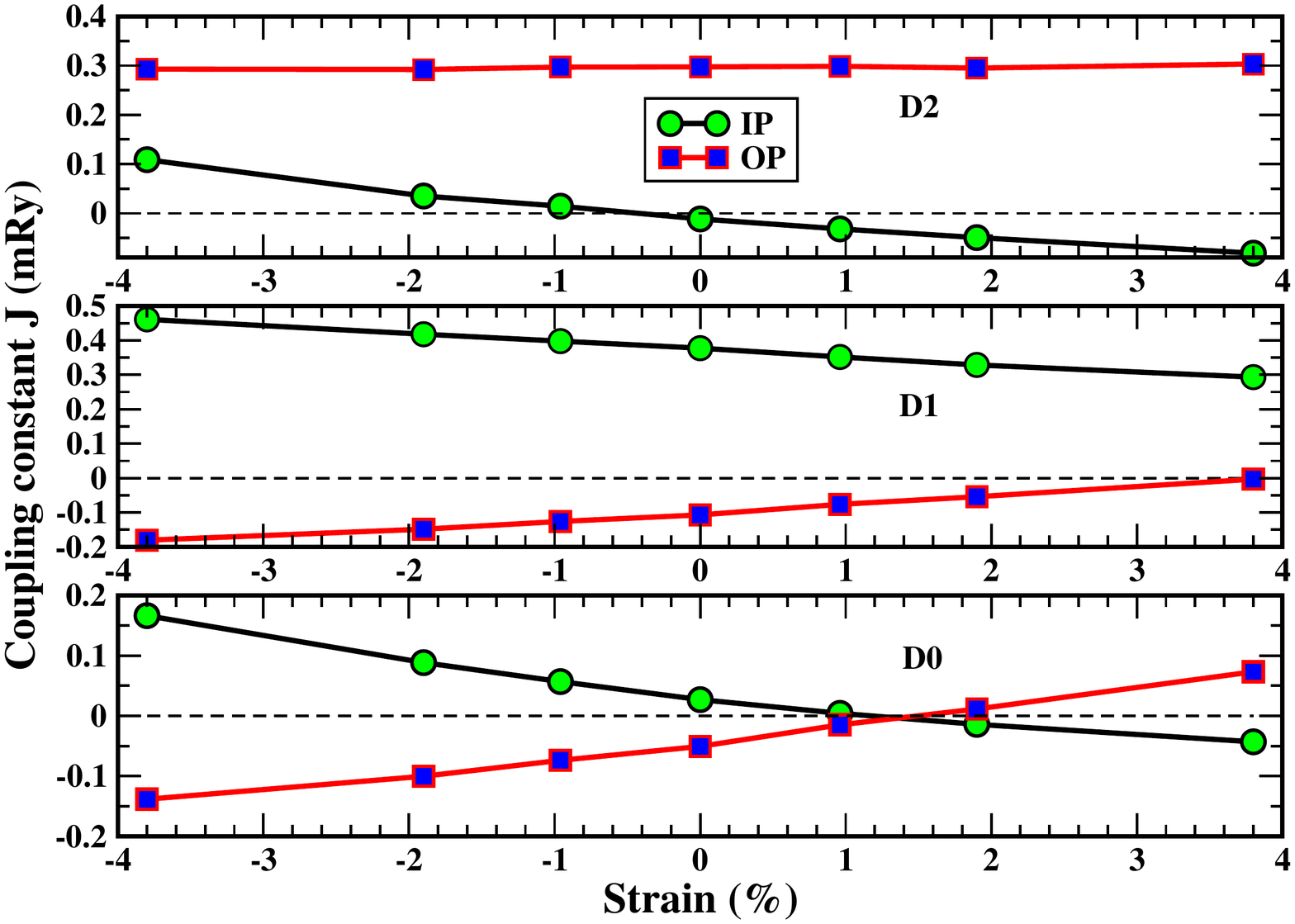}
\label{fig2:subfig1}
}
\subfigure[]
{
\includegraphics[trim= 35 50 35 80, clip, width=0.45\textwidth]{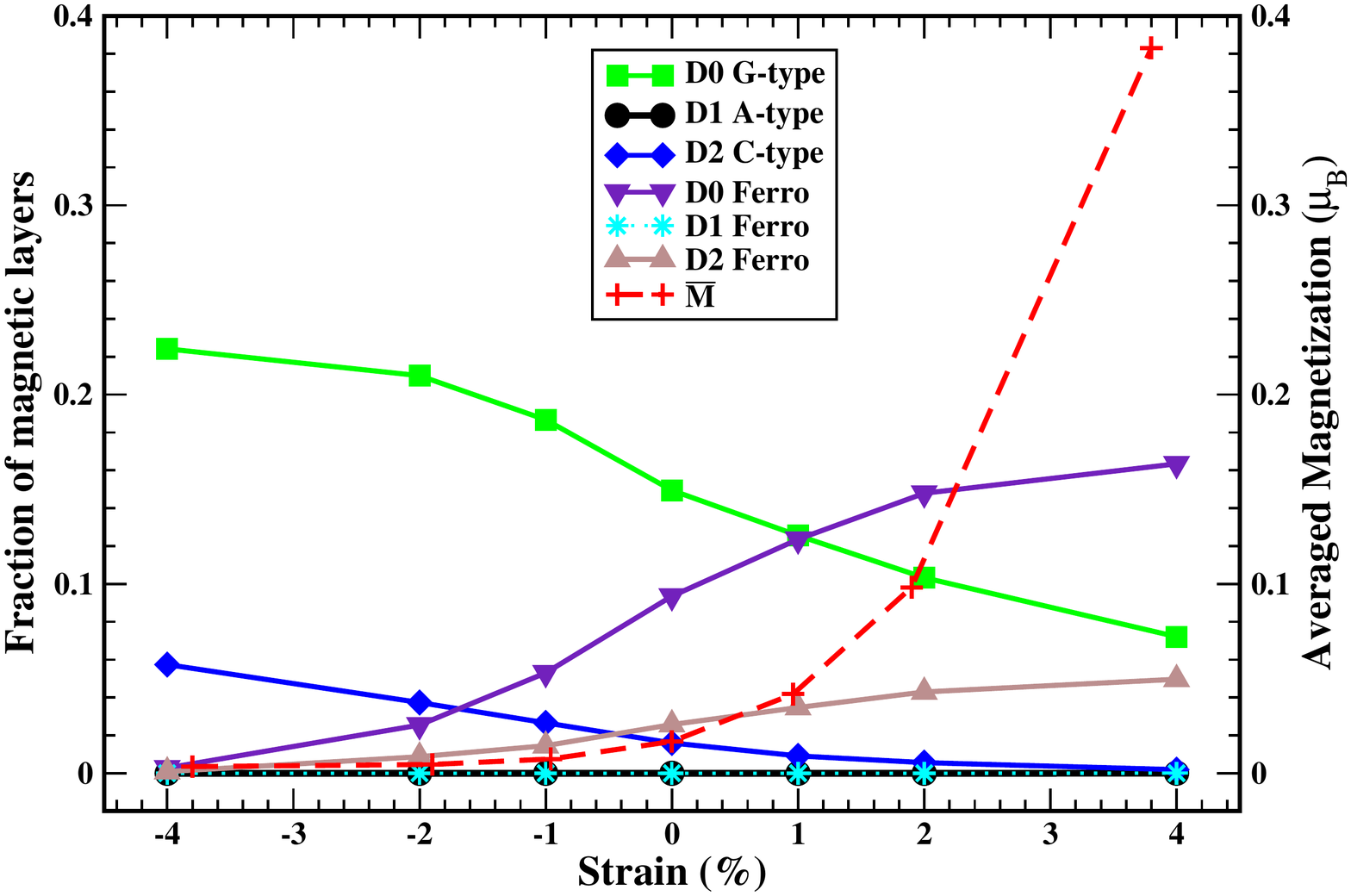}
\label{fig2:subfig2}
}
\caption{Strain dependence of (a) the in-plane (IP, $J_\parallel$) and out-of-plane (OP, $J_{\perp}$) coupling constants in the D0, D1 and D2 structures, and (b) the fraction of magnetic domains obtained from Monte Carlo simulations and the magnetisation averaged over all three structures, $\overline{M}$. (Labels are indicated in the respective figures.) }
\end{figure}

To explain the strain-dependence of the magnetic moment we mapped our 20-atom supercell to a spin-lattice model with only the magnetic ions. The unit cell has 2 Fe/Cr layers, labeled I and II, and each layer having two ($B$-) sites, labeled $\alpha$ and $\beta$. The following Heisenberg hamiltonian describes the spin-spin interactions in the lattice            
\begin{align}
 H &= J_\parallel\sum_{\substack{i,j,k}} \bigl[\vec{S}^{I,\alpha}_{i,j,k}\cdot \{\vec{S}^{I,\beta}_{i,j,k}+\vec{S}^{I,\beta}_{i-1,j,k}+\vec{S}^{I,\beta}_{i,j-1,k}+\vec{S}^{I,\beta}_{i-1,j-1,k} \} \notag \\
   &\qquad + \vec{S}^{II,\alpha}_{i,j,k}\cdot \{ \vec{S}^{II,\beta}_{i,j,k}+ \vec{S}^{II,\beta}_{i-1,j,k}+\vec{S}^{II,    \beta}_{i,j-1,k}+\vec{S}^{II,\beta}_{i-1,j-1,k} \} \bigr] \notag \\
   &\qquad + J_\perp\sum_{\substack{i,j,k}} \bigl[\vec{S}^{I,\alpha}_{i,j,k}\cdot \{ \vec{S}^{II,\alpha}_{i,j,k-1}+\vec{S}^{II,\alpha}_{i,j,k+1}\} \notag \\
   &\qquad +\vec{S}^{I,\beta}_{i,j,k}\cdot \{ \vec{S}^{II,\beta}_{i,j,k-1}+\vec{S}^{II,\beta}_{i,j,k+1}\} \bigr ]
   \label{hamilt}
\end{align}
where $J_\|$ and $J_\perp$ are the in-plane and out-of-plane coupling constants, respectively; $i,j,k$ are site indices and $\vec{S}$ are spin vectors. Given the large magnitudes of the spin we are dealing with ($\sigma^{Fe}=5/2$ and $\sigma^{Cr}=3/2$) we treat the classical version of this hamiltonian choosing the spin vectors to be parallel to the $z$-axis. Using Eq.~\ref{hamilt} for different magnetically ordered supercells it is straightforward extract the values of $J_\|$ and $J_\perp$ in the D0, D1 and D2 structures in terms of the corresponding supercell total energies. The variation of these coupling constants with epitaxial strain is plotted in Fig.\ref{fig2:subfig1} for the three structures.

 In D0 (bottom panel) $J_\|$ is positive in the strain window $-3.8\%$ to $1\%$, indicating an in-plane AFM coupling between Fe and Cr. Simultaneously, in the same strain window, $J_\perp$ is negative resulting in an FM coupling between the out-of-plane Fe-Cr pair. Thus, the ground state becomes C-type AFM in D0 for compressive strains. Beyond 1\% the coupling constants change signs, i.e the $J_\perp$ become positive giving an AFM coupling and the $J_\|$ becomes negative which in turn leads to an in-plane FM coupling between Fe-Cr pair. As a result of which the D0 system undergoes a C-type to A-type AFM transition as seen in the phase diagram above.
         
The middle panel of Fig.\ref{fig2:subfig1} shows the variation of the coupling constants in the D1 structure. Here, $J_\|$ remains positive over all the strain maintaining an AFM coupling between in-plane Fe-Fe pair as required by the Goodenough-Kanamori rules (GK)~\cite{good}, while at the same time $J_\perp$ remains completely dominating in the negative region over all kind of strain. Hence, an FM coupling between the out-of-plane Fe-Cr pair comes into play. Due to the above coupling the ground state structure is a purely C-type under compressive as well as tensile strain region.

The top panel of Fig.\ref{fig2:subfig1} is for D2 where the in-plane order is AFM from -3.8\% upto -1\% strain ($J_\| > 0$) and FM beyond that strain ($J_\| < 0$). However, $J_\perp$  remains positive over the whole strain region yielding an out-of-plane AFM order. Therefore, there is a G-type to A-type magnetic ground state transition in the phase diagram (Fig.\ref{fig1:subfig2}) for D2. Since the magnititude of $J_\parallel$ of D1 remains larger than D0 and D2 the  ground state is a C-type D1-ordered structure. 

In general, the superexchange interaction between like ions (Fe-Fe and Cr-Cr) was seen to be AFM under all strains, as dictated by GK. However, the results above indicate that in the case of Fe-Cr superexchange pairs there is a strong dependence on the strain, and more directly, on the Fe-O-Cr bond angle. It was observed that (Fig.\ref{figureS2} in SI) for Fe-O-Cr bond angles above $\approx149^\circ$ the in-plane coupling is FM while for lower angles an AFM coupling is preferred. The out-of-plane coupling also changes sign but at $\approx147^\circ$. Similar sign reversal of the Fe-Cr coupling with increasing bond angles has been reported for Fe-doped rare-earth orthochromites earlier from M\"{o}ssbauer effect measurements~\cite{moskvin}. An ambiguity in applying GK to $d^3-d^5$ pairs arises because of the differing angular dependences of the dominant kinetic and potential exchange terms in the superexchange. In this case these dominant pathways are Fe($t_{2g}$)-O($p_\pi$)-Cr($t_{2g}$) for kinetic (AFM) and Fe($e_{g}$)-O($p_\sigma$)-Cr($t_{2g}$) for potential (FM) exchange. The latter is generally weaker in strength than the AFM pathway~\cite{LFCO1}. However, given the difference in strength and orientation of the corresponding orbital overlaps ($\sigma$ {\it vs.} $\pi$) in the two cases there is a strong dependence of the dominant pathway on the geometry of the bond. Since compressive strain forces lower bond angles ($\lessapprox149^\circ$) for in-plane pairs and larger ones for out-of-plane pairs generally the AFM dominates in-plane and FM dominates out-of-plane (justifying the stability of C-type AFM). Since the FM interactions are rather weak the lowest excitation under compressive strains often turns out to be out-of-plane AFM (G-type or A-type). For tensile strains the situation is reversed yielding a preference towards A-type AFM ground-state in both D0 and D2 with in-plane Fe-Cr interactions. 

The phase diagram in Fig.\ref{fig1:subfig2} indicates that magnetic excitations are thermally accessible in all structures. This motivated us to study the finite temperature behaviour of the thin films. Using Monte Carlo (MC) simulations on a 60-layer thick film we sampled the configurations allowed by the hamiltonian above at 300K at each strain. For each type of arrangement of the cations we monitored the population of the domains yielding non-zero magnetic moment. This is plotted as the fraction of the layers belonging to a domain versus the strain in Fig.\ref{fig2:subfig2} along with the averaged magnetisation ($\overline{M}$). The drastic reduction in the magnitude of the magnetisation in the coherent strain window observed in experiments is reproduced. Note that the estimated magnitudes are an upper bound to the magnetisation assuming alignment among the magnetic domains. Nevertheless, the correlation between the trend seen in Fig.\ref{fig2:subfig2} and the phase diagram in Fig.\ref{fig1:subfig2} strongly indicate that our finite temperature predictions consistent with our first-principles calculations.
   
Our results caution against the direct extrapolation of bulk results to epitaxially grown thin films. Thin-film BFCO may not have the G-type AFM ground-state expected in its bulk form. In contrast, under all coherent epitaxial strains BFCO favours a C-type AFM magnetic ordering. Furthermore, the double-perovskite structure is unstable under all coherent strains towards cation disordering. In all cases considered the ground-state has zero magnetisation and magnetic moment arises from thermal excitations to the G-type AFM and FM orders. We confirm this trend using MC simulations which also yield a value for the magnetisation which is significantly lower that the expected bulk value. Our predictions match quite well with similar observations made in recent experiments~\cite{khare,neh1}. In fact, a recent work~\footnote{D. S. Rana (private communication)} measuring XMCD and chemically sensitive hysteresis loops on BFCO thin film samples have indeed confirmed the presence of both very low magnetic moments as well as cation disorder. By analysing the dependence of the exchange coupling constants on the strain and bond geometries we conclude that the strength and sign of the Fe-Cr interaction is tunable by strain suggesting interesting possibilities for BFCO and other materials of this kind. 

PLD growth conditions could affect the proportions to which the film incorporates the various cation orderings during
formation. This would result in differences in observed magnetisation dependences. However, our work suggests that intrinsically BFCO displays a very interesting evolution of
magnetic and chemical order with epitaxial strain. We hope that more consequences to this interesting phase diagram will be experimentally probed to reveal other 
exciting phenomena in the material. 
\acknowledgements{The authors would like to thank Dr. D. S. Rana  and Dr. R. S. Singh at IISER Bhopal, Dr. S. Lal at IISER Kolkata and Dr. M. Coccocioni at EPFL, Lausanne for 
helpful discussions. The authors gratefully acknowledge IISER Bhopal for computational resources and funding. PCR would like to acknowledge CSIR for funding 
through the JRF programme.}

\newpage

\begin{center}
{\Large \bf Supplementary Information}
\end{center}

\section{Details of DFT calculations}
The spin-polarised DFT calculations in this work employed GGA+$U$ approach using the Perdew-Burke-Ernzerhof~\cite{pbe1,pbe2} exchange-correlation functional. The  simplified version of DFT+$U$ functional given by Coccocioni {\it et al}~\cite{matteo,wentzco} and as implemented in the Quantum-ESPRESSO code~\cite{paulo} was used. We chose a value of 3.0 eV for the Hubbard parameter $U$  which reproduced well the band structure obtained in previous LSDA+$U$ calculations on BFCO~\cite{baettig,Ghosez}. But, the Hubbard $U$ parameter can, in principle, be calculated {\it ab initio} using the linear response formalism of Ref.~\cite{matteo,wentzco}. We found the $U$ values to be of 4.25 eV (Fe) and 3.07 eV (Cr) in BFCO. However, these neither altered the band structure significantly nor affected the phase diagram for the strained thin films beyond the accuracy of our calculations. So, to stay consistent with literature, we proceeded with $U=3$ eV. We used a $8\times 8\times 8$ Monkhorst-Pack k-point mesh for Brillouin zone integration and a plane-wave basis with a 58 Ry kinetic-energy cutoff along with a 600 Ry cutoff for the charge-density. Ionic cores were modelled by ultra-soft pseudopotentials. Structures were relaxed until the forces are less than 13 meV/$\angs$.

\section{Strain-dependence of out-of-plane lattice parameter}
\begin{figure}[t!]
\begin{center}
\subfigure[]
{
\includegraphics[width=0.45\textwidth]{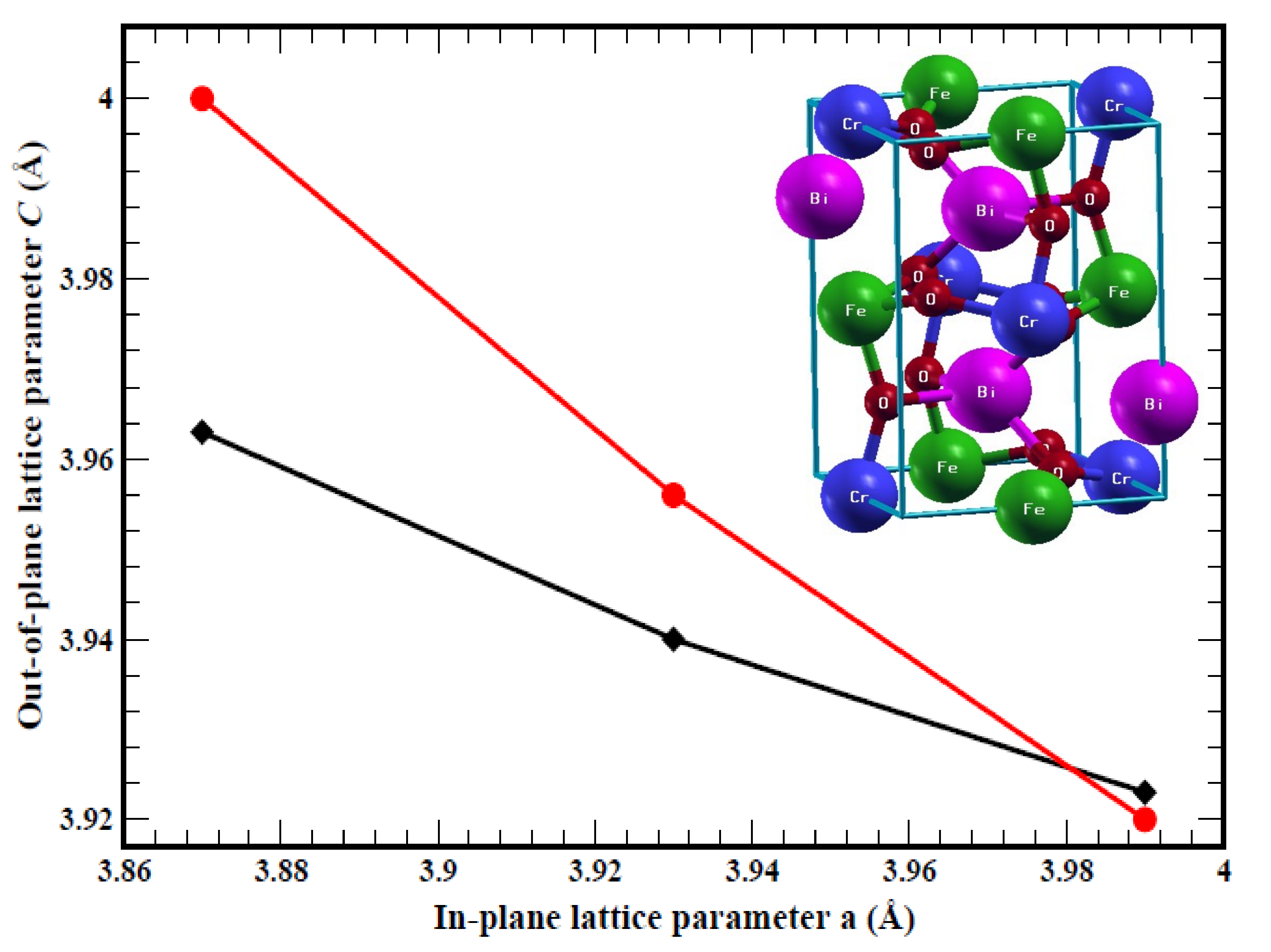}
\label{figureS1a}
}
\subfigure[]
{
\includegraphics[width=0.45\textwidth]{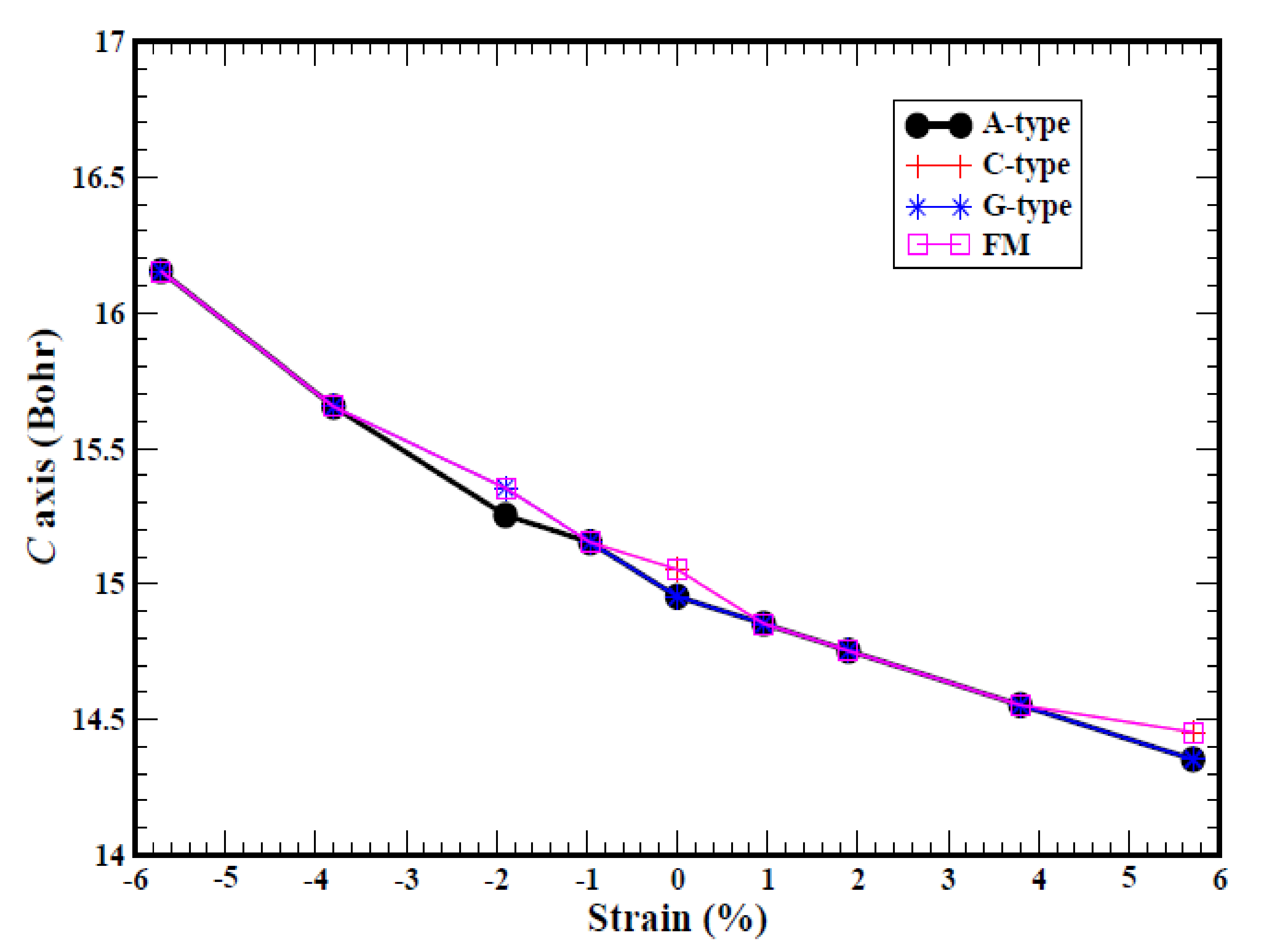}
\label{figureS1b}
}
\caption{Variation of out-of-plane lattice parameter with the in-plane lattice parameter. The curve with circlular symbols represents the GGA+U calculated 
out-of-plane and the line with the square points denotes the experimental one.The inset double perovskite structure is the unit cell used for the calculations}
\end{center}
\end{figure}

The calculated $c_{min}(\epsilon)$ versus in-plane lattice parameter (a) along with the experimental result Khare {\it et al}~\cite{khare} is plotted in Fig.~\ref{figureS1a}
within the coherently strained region. At the compressive strain the calculated out-of-palne lattice parameter ($c_{min}(\epsilon)$) is 1.01{\%} larger than the
experimental value and at tensile strain it underestimates 0.56{\%} to its corresponding experimental value but follows the experimental observed trend. Similar errors have also been reported in previous GGA studies on 
similar materials. See, for instance, the work by Dobin {\it et al}~\cite{dobin}. The inset in Fig.~\ref{figureS1a} is the structure of the BFCO thin film unit cell where the B site cations
are maintained alternatively to get the required magnetic ordering. Fig.~\ref{figureS1b} shows the strain dependence of $c_{min}$ for all magnetic types considered for the double perovskite (D0) structure. There is very little effect of
the change in magnetic order on the lattice parameters.

\section{Strain-dependence of the ${\rm Fe-O-Cr}$ bond angle}

\begin{figure}[t!]
\begin{center}
\subfigure[]
{
\includegraphics[width=0.45\textwidth]{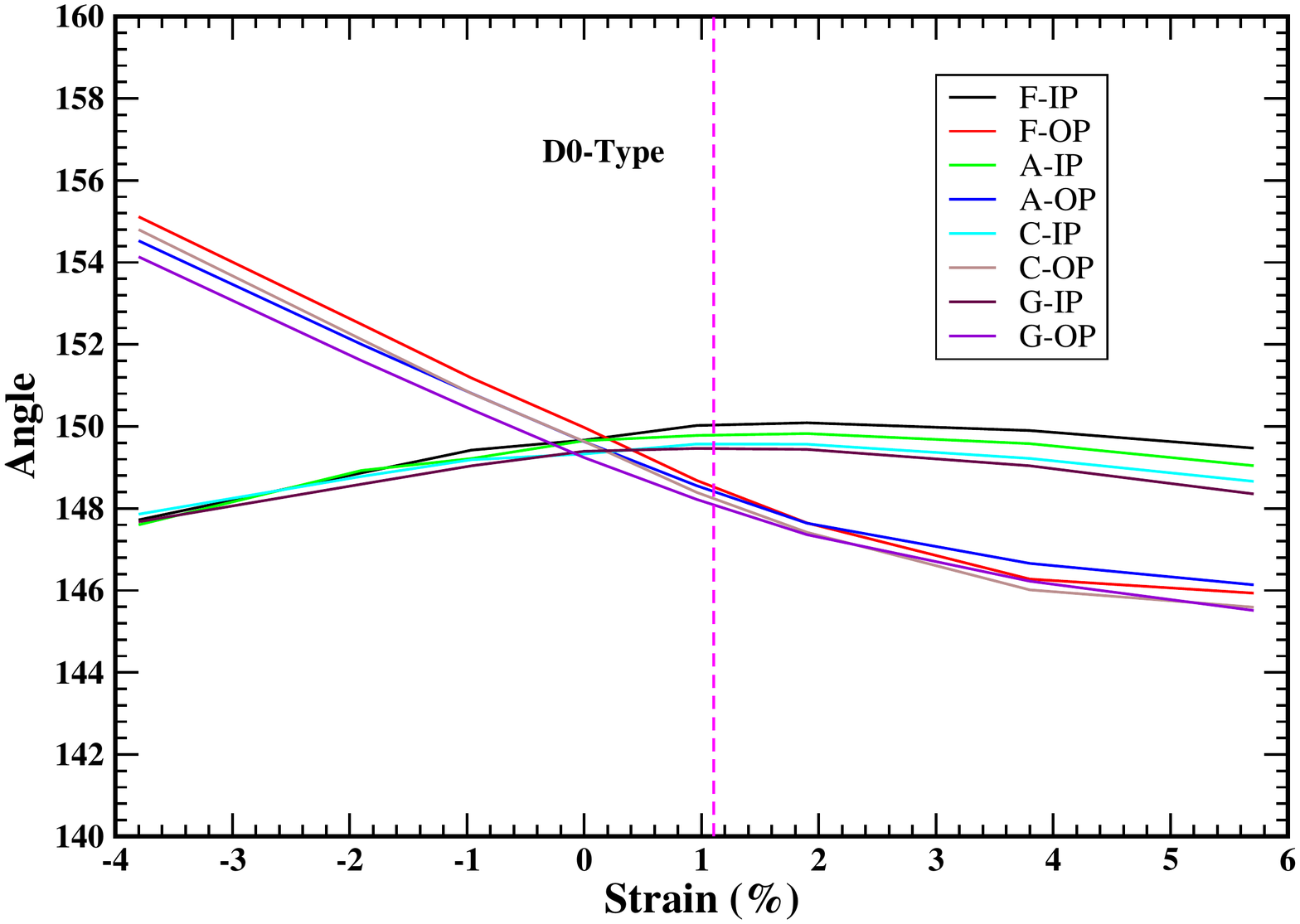}
}
\subfigure[]
{
\includegraphics[width=0.45\textwidth]{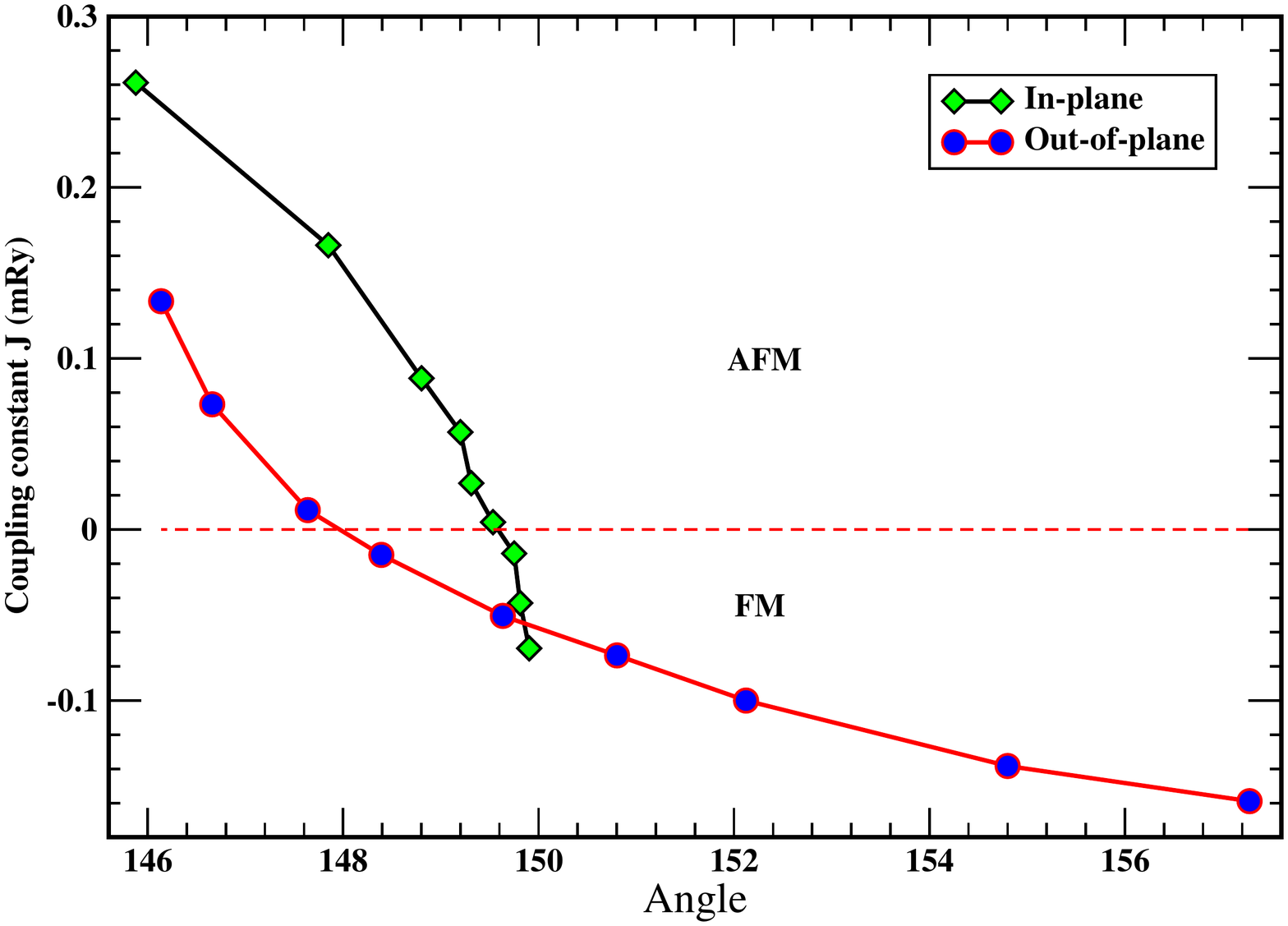}
}
\caption{(a) Variation of out-of-plane (OP) and in-plane (IP) Fe-O-Cr angle under epitaxial strain for different types of magnetic orders in D0 structure. The vertical line marks the transition strain where the film goes from C-type to A-type AFM order, and (b) Variation of OP an IP coupling constants with the corresponding Fe-O-Cr bond angle. Note that the direction of increasing strain is along the direction of increasing IP angle and decreasing OP angle.}
\label{figureS2}
\end{center}
\end{figure}

The Fe-O-Cr bond angle was monitored at every strain for all magnetic orders. Fig.~\ref{figureS2}(a) shows this dependence for the D0 structure. As the ground-state changes from C-type to A-type AFM (around $1\%$ strain) the
in-plane angle increases to $\sim150^\circ$ while the out-of-plane angle is lowered to $\sim146^\circ$.

Strain dependence of bond-angles are also shown for D1 and D2. In D1 the in-plane and in D2 the  out-of-plane angles are for Fe-O-Fe or Cr-O-Cr, i.e. between like cations. These do not undergo any change in sign with strain and remain AFM consistent with the Goodneough-Kanamori rules. 

\begin{figure}[t!]
\begin{center}
\subfigure[]
{
\includegraphics[width=0.45\textwidth]{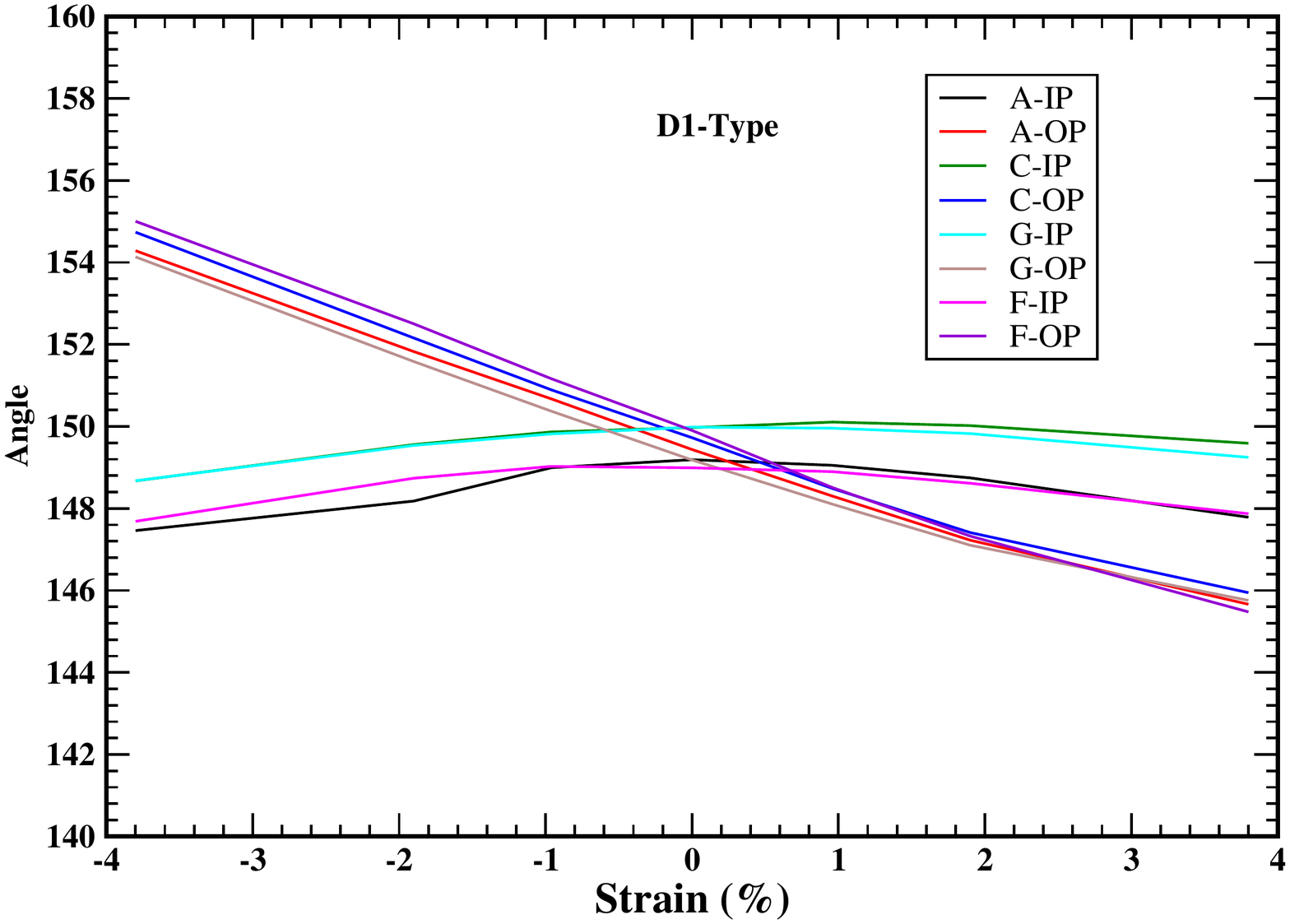}
}
\subfigure[]
{
\includegraphics[width=0.45\textwidth]{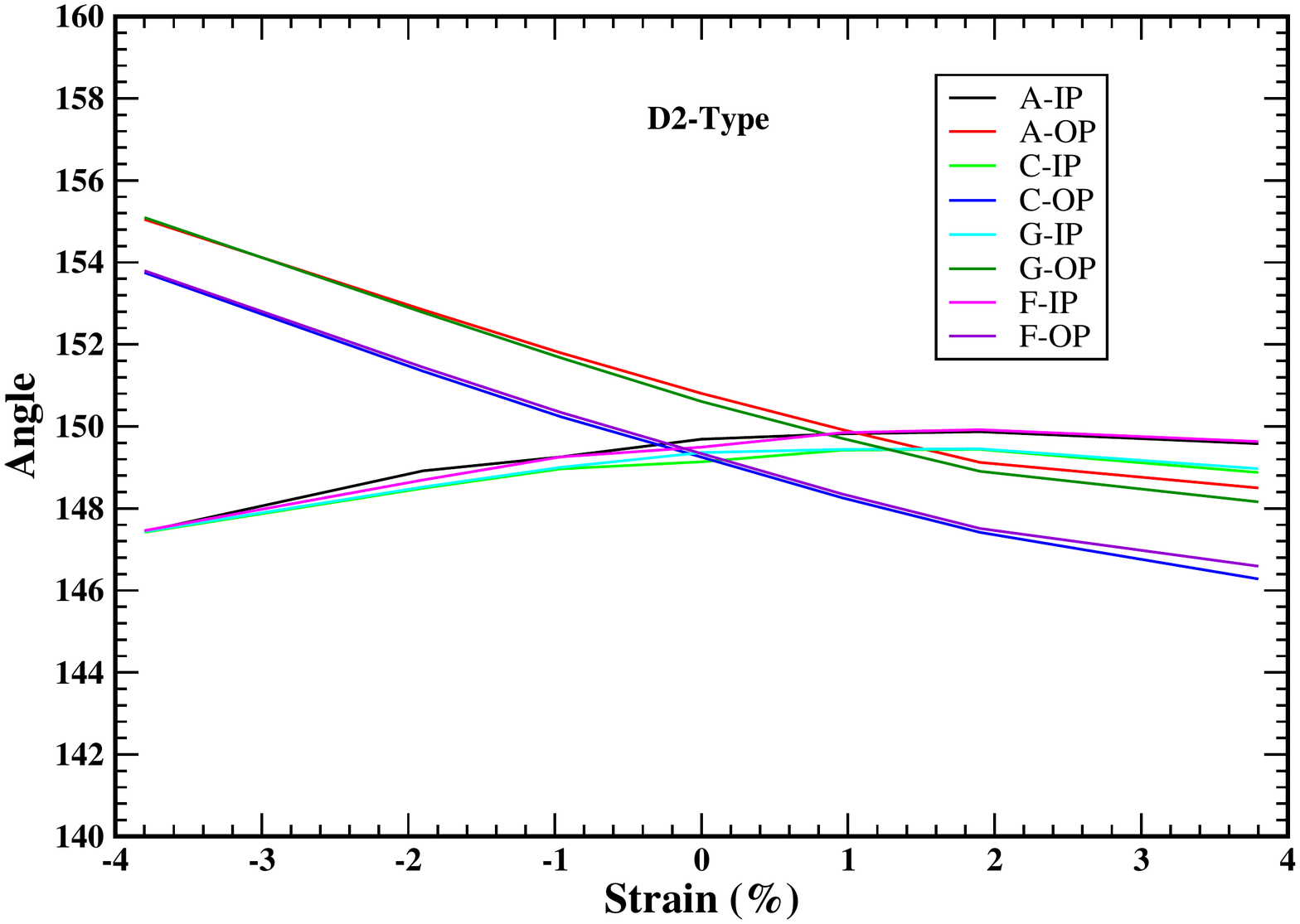}
}
\caption{Variation of out-of-plane (Fe-O-Cr) and in-plane (Fe-O-Fe or Cr-O-Cr) angle under epitaxial strain for different types of magnetic orders in (a) D1 structure, and (b) D2 structure.}
\label{figureS3}
\end{center}
\end{figure}

\section{Monte Carlo simulations}
The following Heisenberg hamiltonian describes the spin-spin interactions in the lattice            
\begin{align}
 H &= J_\parallel\sum_{\substack{i,j,k}} \bigl[\vec{S}^{I,\alpha}_{i,j,k}\cdot \{\vec{S}^{I,\beta}_{i,j,k}+\vec{S}^{I,\beta}_{i-1,j,k}+\vec{S}^{I,\beta}_{i,j-1,k}+\vec{S}^{I,\beta}_{i-1,j-1,k} \} \notag \\
   &\qquad + \vec{S}^{II,\alpha}_{i,j,k}\cdot \{ \vec{S}^{II,\beta}_{i,j,k}+ \vec{S}^{II,\beta}_{i-1,j,k}+\vec{S}^{II,    \beta}_{i,j-1,k}+\vec{S}^{II,\beta}_{i-1,j-1,k} \} \bigr] \notag \\
   &\qquad + J_\perp\sum_{\substack{i,j,k}} \bigl[\vec{S}^{I,\alpha}_{i,j,k}\cdot \{ \vec{S}^{II,\alpha}_{i,j,k-1}+\vec{S}^{II,\alpha}_{i,j,k+1}\} \notag \\
   &\qquad +\vec{S}^{I,\beta}_{i,j,k}\cdot \{ \vec{S}^{II,\beta}_{i,j,k-1}+\vec{S}^{II,\beta}_{i,j,k+1}\} \bigr ]
   \label{hamilt}
\end{align}
where $J_\|$ and $J_\perp$ are the in-plane and out-of-plane coupling constants, respectively; $i,j,k$ are site indices and $\vec{S}$ are spin vectors. Given the large magnitudes of the spin we are dealing with ($\sigma^{Fe}=5/2$ and $\sigma^{Cr}=3/2$) we treat the classical version of this hamiltonian choosing the spin vectors to be parallel to the $z$-axis. Using Eq.~\ref{hamilt} for different magnetically ordered supercells it is straightforward extract the values of $J_\|$ and $J_\perp$ in the D0, D1 and D2 structures in terms of the corresponding supercell total energies. For instance, for the D0 structure we have
\begin{eqnarray}
J_\parallel=\frac{E_{A}-E_{G}}{60} ,~J_\perp=\frac{E_{C}-E_{G}}{30}
\end{eqnarray}
Similar expressions can be obtained for the coupling constants in the D1 and D2 structures as well. 

Using Monte Carlo (MC) simulations on a 60-layer thick film we sampled the configurations allowed by the hamiltonian above at 300K at each strain. Furthermore, the  MC moves were restricted to those that flip spins of an entire sub-lattice in a given layer. This was done in order to accelerate the sampling since we were primarily interested in the population of domains of each magnetic order in the thin-film. For each type of arrangement of the cations we monitored the population of the domains yielding non-zero magnetic moment. The fraction of the layers belonging to a domain versus the strain is plotted in the main text in Fig.2(b).   

The average magnetic moment can be estimated by the sum of the contributions of the magnetic domains from each structure weighted by the probability of the thin-film occurring in a given structure 
\begin{eqnarray}
\overline{M}(\epsilon) = \sum\limits_{\nu} w^\nu(\epsilon) \left( \sum\limits_X n^\nu_X(\epsilon)~m^\nu_X\right)  
\end{eqnarray}
where $n_X^\nu(\epsilon)$ is the fraction of $X$-type magnetic domain in $\nu$-type structure at the strain $\epsilon$ and $m^\nu_X$ is the magnetic moment per formula unit associated with a $X$-type magnetic domain in $\nu$-type structure. $X$ and $\nu$ run over the different magnetic and structure types D0, D1 and D2 (see main text). $w^\nu(\epsilon)$ is the weight associated with the occurrence of the $\nu$-type structure. The latter is approximated as the probability of occurrence of the ground-state magnetic domain in that structure 
\begin{eqnarray}
w^\nu(\epsilon) \approx \frac{\sum\limits_\tau{exp(-\beta E^\nu_\tau)}}{Z};~Z = \sum\limits_\nu \sum\limits_\tau exp(-\beta E^\nu_\tau)
\end{eqnarray}
where $E^\nu_\tau$ refers to the energy in the magnetic domain $\tau$ ($\tau=$ A,C,G \& Ferro) of the $\nu$-type structure. The denominator in the above equation is chosen to ensure normalisation. $\overline{M}$ is also plotted as a function of strain in Fig.2(b) of the main text.

Note that the estimated magnitudes are at most an upper bound to the magnetisation assuming alignment among the magnetic domains. The latter is obviously not necessary since at finite temperature out-of-plane fluctuations may lead to destruction of any order in the thin-film. Also in our choice of MC moves we have ignored in-plane fluctuations of the spins which might destroy even in-plane order and reduce magnetisation further.

\bibliographystyle{apsrev4-1}
\bibliography{reference}

\end{document}